\documentclass[dvips,12pt,aps,prc,reprint,groupedaddress,superscriptaddress,showpacs]{revtex4-1}
\usepackage{natbib,lipsum}
\usepackage{graphicx,subfigure,float,graphics,amssymb}
\usepackage{epsfig}

\begin{document}

\title{Tilted-axis wobbling in odd-mass nuclei}

\author{R. Budaca}
\affiliation{"Horia Hulubei" National Institute for Physics and Nuclear Engineering, Str. Reactorului 30, RO-077125, POB-MG6 Bucharest-M\v{a}gurele, Romania}

\date{\today}

\begin{abstract}
A triaxial rotor Hamiltonian with a rigidly aligned high-$j$ quasiparticle is treated by a time-dependent variational principle, using angular momentum coherent states. The resulting classical energy function have three unique critical points in a space of generalized conjugate coordinates, which can minimize the energy for specific ordering of the inertial parameters and a fixed angular momentum state. Due to the symmetry of the problem, there are only two unique solutions, corresponding to wobbling motion around a principal axis and respectively a tilted-axis. The wobbling frequencies are obtained after a quantization procedure and then used to calculate $E2$ and $M1$ transition probabilities. The analytical results are employed in the study of the wobbling excitations of $^{135}$Pr nucleus, which is found to undergo a transition from low angular momentum transverse wobbling around a principal axis toward a tilted-axis wobbling at higher angular momentum.
\end{abstract}

\pacs{21.10.Re, 23.20.Lv, 27.70.+q}

\maketitle

\section{Introduction}

Atomic nuclei are predominantly spherical or axially symmetric in their ground state. Although more rarely, deviations from axial symmetry are known to occur in certain regions of the nuclide chart \cite{Werner,Moller1,Moller2}. The rigid triaxiality in nuclei, {\it i.e.} when the asymmetry parameter $\gamma$ is frozen to a certain value, is an even more elusive phenomenon. The quantum mechanical properties of the triaxial rigid rotor was firstly used in nuclear physics by Davydov and Filipov \cite{DavFil} who showed that the low lying collective states in some nuclei can be described by the eigenvalues of a rotor Hamiltonian with different values of the moments of inertia (MOI) corresponding to the axes of the intrinsic reference frame.

Even though, triaxiality has important effects on nucleon separation energies \cite{Moller1}, fission barrier height \cite{Moller3,Lu}, fragmentation of the large amplitude collective excitations \cite{GDR1,GDR2}, probability of proton emission \cite{Prot}, it is still difficult to measure directly. Therefore, a lot of effort was directed to the identification of a clear signature for triaxiality such as signature inversion \cite{Beng} or $\gamma$ band staggering \cite{Stachel,Zamfir}. The later also serves as a distinguishing test for rigid and dynamical triaxiality \cite{McC} and was used to propose candidate nuclei for rigid asymmetry \cite{Bon1,Bug1,Bud}.

Stable triaxial shapes are uniquely related to interesting phenomena such as anomalous signature splitting \cite{Hama1}, chiral symmetry breaking \cite{FrauMeng}, and the wobbling excitations \cite{BM}, whose observation is tantamount to the identification of triaxiality. The possibility of wobbling motion at high spin states was discussed first time for even-even nuclei \cite{BM}. As the occurrence of rigid triaxiality in even-even nuclei became more improbable, studies were directed to odd-mass nuclei, where the alignment of the odd particle angular momentum was supposed to facilitate the emergence of a rigid triaxial core. This was first suggested for the Triaxial Strongly Deformed (TSD) bands of $^{163,165}$Lu based on an aligned $i_{13/2}$ proton \cite{Schnack}. Later, {\O}deg{\aa}rd {\it et al.} \cite{Odegard} showed that two such bands in $^{163}$Lu have similar inertial parameters and degree of single-particle angular momentum alignment up to very high spin - a fact specific to bands connected by wobbling excitations. This first confirmation of the wobbling excitations in $^{163}$Lu, was followed by the identification of wobbling bands based on the alignment of the same proton orbital in other neighboring nuclei $^{161}$Lu \cite{161}, $^{165}$Lu \cite{165}, $^{167}$Lu \cite{167} and $^{167}$Ta \cite{Ta}. Recently, the wobbling mode with an aligned $h_{11/2}$ proton has been observed in the odd-even $^{135}$Pr nucleus \cite{Matta,Garg}.

The Bohr-Mottelson (simple) wobbling frequency predicted for even-even systems \cite{BM} is still a good starting point and a useful reference for the study of wobbling excitations in odd-$A$ nuclei \cite{Shimizu1,Hama2,Frau}. Its adaptation to the presence of an aligned odd particle was realized only recently in a semiclassical description of a triaxial rigid rotor Hamiltonian with alignment \cite{Frau}. The semiclassical approach to general rotor Hamiltonians has the advantage of keeping close contact with the classical features of the system's dynamics \cite{Onishi,Gheorghe,RaBu,Chen,Chen2,Raduta}. The result of Ref.\cite{Frau} explained the origin of the observed decrease in the wobbling excitation as a function of total angular momentum in terms of a so called transverse wobbling which was proposed originally in Ref.\cite{Shimizut}. It is generated by the alignment of the quasiparticle angular momentum perpendicular to the axis with the largest MOI. All wobbling bands observed until now in odd-$A$ nuclei exhibit a transversal type of wobbling. The last entry into the experimentally observed wobbling modes, $^{135}$Pr, is the only case which exhibits a termination of the transversal wobbling. This feature is related to the angular momentum dependence of the existence condition for the transversal wobbling. The critical spin where it terminates marks the transition from wobbling motion around a principal axis to one around a tilted-axis \cite{Heiss,Matsu,Frau,Chen3}.

In this study, one will show that through a rigorous semiclassical treatment of a triaxial rigid rotor Hamiltonian with particle alignment, one obtains besides the longitudinal and transversal wobbling regimes also some complementary modes corresponding to a tilted-axis wobbling motion. The transition between principal and tilted-axis wobbling is used to describe the anomaly in the wobbling excitation energy around $I=29/2$ in the $^{135}$Pr nucleus.

\renewcommand{\theequation}{2.\arabic{equation}}
\section{Theoretical framework}
\subsection{Semiclassical description}

For the description of the interaction between single-particle and collective angular momenta the following particle-rotor Hamiltonian is employed:
\begin{equation}
H=H_{R}+H_{sp},
\end{equation}
where $H_{R}=\sum_{k=1,2,3}A_{k}(\hat{I}_{k}-\hat{j}_{k})^{2}$ is the triaxial rotor Hamiltonian associated to the core angular momentum $\vec{R}=\vec{I}-\vec{j}$ and defined by the inertial parameters $A_{k}$. The later are related to the MOI by $A_{k}=1/(2\mathcal{J}_{k})$. The single-particle contribution to the total Hamiltonian is
\begin{eqnarray}
H_{sp}&=&\frac{V}{j(j+1)}\left\{\left[3\hat{j}_{3}^{2}-j(j+1)\right]\cos{\gamma}\right.\nonumber\\
&&\left.-\sqrt{3}(\hat{j}_{1}^{2}-\hat{j}_{2}^{2})\sin{\gamma}\right\},
\end{eqnarray}
where $\gamma$ is the asymmetry parameter, which also defines the ratios between MOI. In case of one fully aligned particle with an alignment $\hat{j}_{1}\approx j\equiv const.$, the relevant part of the particle-rotor coupling Hamiltonian may be reduced to:
\begin{eqnarray}
H_{align}&=&A_{1}(\hat{I}_{1}-j)^{2}+A_{2}\hat{I}_{2}^{2}+A_{3}\hat{I}_{3}^{2}+const.\nonumber\\
&=&H'_{R}-2A_{1}j\hat{I}_{1}+const.,
\end{eqnarray}
where $\displaystyle H'_{R}=A_{1}\hat{I}_{1}^{2}+A_{2}\hat{I}_{2}^{2}+A_{3}\hat{I}_{3}^{2}$ is a pure rotor Hamiltonian for the total angular momentum $I$. Thus, the Hamiltonian to be treated is:
\begin{equation}
H_{align}=A_{1}\hat{I}_{1}^{2}+A_{2}\hat{I}_{2}^{2}+A_{3}\hat{I}_{3}^{2}-2A_{1}j\hat{I}_{1}.
\label{Hal}
\end{equation}
Because of difficulties in treating the full degrees of freedom associated to the above Hamiltonian, it is desirable to describe it by means of only few classical variables which are extracted in such a way as to be associated to some particular dynamics of the quantum system. An example in this sense is the time-dependent Hartree-Fock theory, which is widely used in the study of nuclear structure and dynamics. Such semiclassical approaches rely on a time-dependent variational principle applied to a variational state which is constructed according to the problem under consideration \cite{Kramer}. The variational principle provides the time-dependence of some restricted set of complex variables which parametrize the variational state. Solving then the equations of motion for the complex variables provided by the variational principle one obtains the classical description of the relevant dynamics of the original quantum system. Moreover, if the variational state spans the whole Hilbert space of the quantum system, the classical equations of motion for its complex variables are equivalent to the original quantum eigenvalue problem. In this context, coherent states are perfect trial functions due to their completeness property, while its continuous character brings a natural transition between quantum and classical pictures \cite{Klauder}.

For the purpose of investigating wobbling excitations emerging from the quantum Hamiltonian (\ref{Hal}), the later is treated within the variation principle
\begin{equation}
\delta\int_{0}^{t}\langle\psi(z)|H_{align}-\frac{\partial}{\partial{t'}}|\psi(z)\rangle dt'=0.
\label{var}
\end{equation}
The variational state $|\psi(z)\rangle$ is chosen of the form:
\begin{equation}
|\psi(z)\rangle=\mathcal{N}e^{z\hat{I}_{-}}|I,I\rangle,
\label{coh}
\end{equation}
where $z$ is a complex time-dependent variable, $|I,M\rangle$ are the eigenstates of the angular momentum operators $\hat{I}^{2}$ and $\hat{I}_{3}$, while $\mathcal{N}=(1-|z|^{2})^{-I}$ is a factor that assures that the function $|\psi(z)\rangle$ is normalized to unity. The spin coherent states of this type are actually generalizations of the famous Glauber coherent states \cite{Glauber} to arbitrary Lie group structures \cite{Perelomov,Radcliffe,Gilmore}.

The averages on the variational state of the terms involved in the variation (\ref{var}) are calculated using the results of Refs.\cite{Radcliffe,Takeno,RaBu} and have the following expressions:
\begin{eqnarray}
\langle H_{align}\rangle&=&\frac{I}{2}(A_{1}+A_{2})+A_{3}I^{2}+\frac{I(2I-1)}{2(1+zz^{*})^{2}}\nonumber\\
&&\times\left[A_{1}(z+z^{*})^{2}-A_{2}(z-z^{*})^{2}-4A_{3}zz^{*}\right]-\nonumber\\
&&\frac{2A_{1}jI(z+z^{*})}{1+zz^{*}},\\
\left\langle\frac{\partial}{\partial{t}}\right\rangle&=&\frac{I(\dot{z}z^{*}-z\dot{z}^{*})}{1+zz^{*}}.
\end{eqnarray}
$z$ and its complex conjugate counterpart are considered as independent variables. The time dependent variational equation (\ref{var}) offers the following equations of motion for the complex variables $z$ and $z^{*}$:
\begin{equation}
\frac{\partial{\mathcal{H}}}{\partial{z}}=-\frac{2iI\dot{z}^{*}}{(1+zz^{*})^{2}},\,\,\,\,\frac{\partial{\mathcal{H}}}{\partial{z^{*}}}=\frac{2iI\dot{z}}{(1+zz^{*})^{2}},
\end{equation}
where $\mathcal{H}=\langle H_{align}\rangle$ plays now the role of a classical energy function which is also a constant of motion. For simplicity, the complex variable is written in a stereographic representation \cite{Radcliffe,Arecchi}
\begin{equation}
z=\tan{\frac{\theta}{2}}e^{i\varphi},\,\,\,0\leq\theta<\pi,\,\,\,0\leq\varphi<2\pi.
\end{equation}
Within this parametrization, the angular momentum carried by the coherent state is oriented in the direction specified by the two angles of rotation $\theta$ and $\varphi$ \cite{Takeno}. The equations of motion for the new variables are given as:
\begin{equation}
\frac{\partial{\mathcal{H}}}{\partial{\theta}}=-I\sin{\frac{\theta}{2}}\dot{\varphi},\,\,\,\,\frac{\partial{\mathcal{H}}}{\partial{\varphi}}=I\sin{\frac{\theta}{2}}\dot{\theta}.
\end{equation}

The full structure of the classical Hamiltonian system is reproduced if the variables are canonical. This is achieved by the change of variable
\begin{equation}
r=2I\cos{\frac{\theta}{2}},\,\,\,0\leq r\leq 2I.
\end{equation}
With this, the equations of motion acquire the canonical Hamilton form
\begin{equation}
\frac{\partial{\mathcal{H}}}{\partial{r}}=\dot{\varphi},\,\,\,\,\frac{\partial{\mathcal{H}}}{\partial{\varphi}}=-\dot{r},
\end{equation}
and one can distinguish now the role of each conjugate generalized coordinates. Thus, $\varphi$ is the generalized coordinate, while $r$ is the generalized momentum. This distinction will become handy for the quantization procedure. The classical energy function have the following expression in terms of the canonical variables:
\begin{eqnarray}
\mathcal{H}(r,\varphi)&=&\frac{I}{2}(A_{1}+A_{2})+A_{3}I^{2}+\frac{(2I-1)r(2I-r)}{2I}\nonumber\\
&&\times(A_{1}\cos^{2}{\varphi}+A_{2}\sin^{2}{\varphi}-A_{3})-\nonumber\\
&&2A_{1}j\sqrt{r(2I-r)}\cos{\varphi}.
\label{clase}
\end{eqnarray}
The classical trajectory of the angular momentum vector $\vec{I}$ is a curve in the space of its classical projections $(I_{1},I_{2},I_{3})$ on the principal axes determined by the intersection of the constant energy surfaces provided by the constants of motions. These are the classical energy function and the total angular momentum:
\begin{eqnarray}
\mathcal{H}&=&A_{1}I_{1}^{2}+A_{2}I_{2}^{2}+A_{3}I_{3}^{2}-2A_{1}jI_{1},\\
I^{2}&=&I_{1}^{2}+I_{2}^{2}+I_{3}^{2}.
\end{eqnarray}
In the space of classical components, the first condition is represented by a shifted ellipsoidal surface, while the second is a sphere. The conservation of the total angular momentum can be easily verified by employing the expressions of classical angular momentum components as functions of the canonical variables \cite{Ida,RaBu}:
\begin{eqnarray}
I_{1}&=&\sqrt{r(2I-r)}\cos{\varphi},\nonumber\\
I_{2}&=&\sqrt{r(2I-r)}\sin{\varphi},\label{Icl}\\
I_{3}&=&r-I.\nonumber
\end{eqnarray}
The three coordinates are then reduced to only two, which are taken to be the canonical variables $\varphi$ and $r$. The final purpose is to obtain the energy spectrum. This is usually done by quantizing the period of the classical closed orbits obtained as solutions of the equations of motion \cite{Gheorghe,Onishi,Frau}. The procedure is similar to the Bohr-Sommerfeld quantization condition. The classical orbits are closed curves in the phase space of the canonical coordinates which are concentrically positioned around the stationary points of the constant energy surface. The stationary points where the time derivatives of the canonical variables vanish, are determined from the critical point condition for the classical energy function:
\begin{equation}
\left(\frac{\partial{\mathcal{H}}}{\partial{r}}\right)_{r_{0},\varphi_{0}}=0,\,\,\,\,\left(\frac{\partial{\mathcal{H}}}{\partial{\varphi}}\right)_{r_{0},\varphi_{0}}=0.
\label{min}
\end{equation}
If one considers $\sqrt{r(2I-r)}>0$, there are three stationary points of the classical energy function $\mathcal{H}(r,\varphi)$ which are listed in Table \ref{tab1} using the following simplifying notation
\begin{equation}
\cos{\alpha_{2,3}}=\frac{2A_{1}j}{(2I-1)(A_{1}-A_{2,3})}.
\label{alpha}
\end{equation}
The indexing of the $\alpha$ angle is related to the axes perpendicular to the first axis with the single-particle angular momentum alignment. Its physical meaning will become clear when the classical motion of the system will be discussed.

Only the stationary points which minimize the classical energy are of interest. The conditions in which all three critical points become minima are determined by studying the corresponding Hessian matrix. It comes down to having a positive determinant of the Hessian matrix, and one of its diagonal minors. The domain of existence for the resulted minima are indicated in Table \ref{tab1} using the following angular momentum weighting factor for MOI
\begin{equation}
S_{Ij}=\frac{2I-1-2j}{2I-1}.
\end{equation}
No prior ordering relation for the inertial parameters $A_{k}(k=1,2,3)$ was considered. The conditions for existence of the solutions 2 and 3 as minima, imply that $\cos{\alpha_{2,3}}>0$ and implicitly $\alpha<\pi/2$.

\setlength{\tabcolsep}{5.5pt}
\begin{table*}[ht!]
\caption{Critical points of the classical energy function (\ref{clase}) with corresponding restrictions which make them minima. The listed values of the classical angular momentum components in the minimum points uniquely identify the distinct wobbling modes. Last column of the table shows the set of three Euler angles which transforms the original reference frame to the one aligned to the average direction of the angular momentum vector.}
\label{tab1}
\begin{center}
\begin{tabular}{cccccccc}
\hline\hline\noalign{\smallskip}
$i$ & $(r_{i},\varphi_{i})$& Conditions                          & $I_{1}^{cl}$ & $I_{2}^{cl}$ & $I_{3}^{cl}$ &$(\psi,\theta,\phi)$\\
\noalign{\smallskip}\hline\noalign{\smallskip}
1   & $\left(I,0\right)$                                                                     & $S_{Ij}A_{1}<A_{2}<A_{3}$                 & $I$            &  0          & 0 & $(\pi/2,\pi/2,\pi)$ \\
    &                                                                             & $S_{Ij}A_{1}<A_{3}<A_{2}$                 &                &             &  &\\
\noalign{\smallskip}\hline\noalign{\smallskip}
2   & $\left(\sqrt{r_{2}(2I-r_{2})}=I,\alpha_{2}\right)$& $A_{2}<A_{3}<A_{1}S_{Ij}$       & $I\cos{\alpha_{2}}$&  $I\sin{\alpha_{2}}$& 0& $(\pi/2+\alpha_{2},\pi/2,\pi)$\\
    &                                                                             & $A_{2}<S_{Ij}A_{1}<A_{3}$      &                &             & &  \\
\noalign{\smallskip}\hline\noalign{\smallskip}
3   & $\left(\sqrt{r_{3}(2I-r_{3})}=I\cos{\alpha_{3}},0\right)$& $A_{3}<A_{2}<A_{1}S_{Ij}$           & $I\cos{\alpha_{3}}$&  0          & $I\sin{\alpha_{3}}$ & $(\pi/2,\pi/2+\alpha_{3},\pi)$\\
    &                                                                             & $A_{3}<A_{1}S_{Ij}<A_{2}$      &                &             &  & \\
\noalign{\smallskip}\hline\hline
\end{tabular}
\end{center}
\vspace{-0.5cm}
\end{table*}

\subsection{Wobbling excitation energies}

Instead of quantizing the classical orbits, one can exploit the canonicity of the two variables and quantize directly the classical energy function by means of the correspondence principle. Unfortunately, the classical energy function contains mixed terms in generalized coordinate and momentum. Although one can symmetrize such products, the quantization procedure will lose some of its reliability. To avoid that, one first expand the energy function around its minima $(r_{i},\varphi_{i}$) with $i=1,2,3$ and truncate the series at the second order:
\begin{eqnarray}
\mathcal{H}_{i}(r,\varphi)&=&\mathcal{H}(r_{i},\varphi_{i})+\frac{1}{2}\left(\frac{\partial^{2}{\mathcal{H}}}{\partial{r^{2}}}\right)_{r_{i},\varphi_{i}}\tilde{r}_{i}^{2}\nonumber\\
&&+\frac{1}{2}\left(\frac{\partial^{2}{\mathcal{H}}}{\partial{\varphi^{2}}}\right)_{r_{i},\varphi_{i}}\tilde{\varphi}_{i}^{2},
\end{eqnarray}
where $\tilde{r}_{i}=r-r_{i}$ and $\tilde{\varphi}_{i}=\varphi-\varphi_{i}$. In this way, the energy function will acquire the form of a classical oscillator function. The fact that positiveness of the oscillator parameters, {\it i.e.} mass and string constant, is implicitly satisfied, comes from the condition that the critical points to be minima. The classical trajectories become unstable as they depart from a critical point, where their quantization becomes problematic \cite{Gheorghe,Frau}. Therefore, the harmonic approximation of the classical energy function is consistent with the extent of the phase space corresponding to fully quantizable trajectories. Quantizing the resulted oscillator functions by replacing the generalized coordinate and momentum with their operator counterparts, one arrives at the following discrete energy spectra:
\begin{widetext}
\begin{eqnarray}
E_{1}(I,n)&=&A_{1}I^{2}+\frac{I}{2}(A_{2}+A_{3})-2A_{1}jI+\omega_{1}(I)\left(n+\frac{1}{2}\right),\label{E1}\\
E_{2}(I,n)&=&A_{2}I^{2}+\frac{I}{2}(A_{1}+A_{3})-A_{1}jI\cos{\alpha_{2}}+\omega_{2}(I)\left(n+\frac{1}{2}\right),\label{E2}\\
E_{3}(I,n)&=&A_{3}I^{2}+\frac{I}{2}(A_{1}+A_{2})-A_{1}jI\cos{\alpha_{3}}+\omega_{3}(I)\left(n+\frac{1}{2}\right),\label{E3}
\end{eqnarray}
where the associated wobbling frequencies are given by:
\begin{eqnarray}
\omega_{1}(I)&=&\sqrt{\left[(2I-1)(A_{3}-A_{1})+2A_{1}j\right]\left[(2I-1)(A_{2}-A_{1})+2A_{1}j\right]},\label{w1}\\
\omega_{2}(I)&=&(2I-1)\sqrt{(A_{3}-A_{2})(A_{1}-A_{2})}\sin{\alpha_{2}},\label{w2}\\
\omega_{3}(I)&=&(2I-1)\sqrt{(A_{2}-A_{3})(A_{1}-A_{3})}\sin{\alpha_{3}}\label{w3}.
\end{eqnarray}
\end{widetext}
The oscillator quanta $n$ are associated here with the wobbling excitations. A few comments regarding the quantal energies are in order. The non-wobbling terms of the quantum energy describe the rotational motion of the system. The dominant term proportional to $I^{2}$ obviously describes the rotation, while the linear term constitute the precession correction. Within this picture, one can see that the rotation-precession motion corresponding to the solution 1, proceed around the first axis, which was chosen from beginning as the alignment axis. The large $I$ limit of the first frequency is just the result obtained by Frauendorf and D\"{o}nau \cite{Frau}, but with axes 1 and 3 interchanged. The  solutions 2 and 3 recover each other when the axes 2 and 3 are interchanged, and correspond to rotations around the second and respectively the third principal axis. The wobbling frequency for these last solutions is given by the simple wobbling estimation of Bohr and Mottelson \cite{BM} tilted with the angle $\alpha_{2,3}$. The dynamical properties of each solution can also be inferred from the values of the classical angular momentum components (\ref{Icl}) in their corresponding minimum points which are listed in Table \ref{tab1}. From this analysis, the tilted nature of the new wobbling modes is more obvious. Moreover, one can see that the angles $\alpha_{2,3}$ actually describe the departure of the average angular momentum vector from the first principal axis.

\subsection{Electromagnetic transitions}

Using the formalism of \cite{BM}, the reduced matrix element of the $E2$ transition operator can be written in the following form:
\begin{equation}
\langle I'n'||\mathcal{M}(E2)||I,n\rangle=\frac{1}{\sqrt{2I+1}}\sqrt{\frac{5}{16\pi}}e\langle n'|m(I,I')|n\rangle,
\end{equation}
where
\begin{eqnarray}
m(I,I')&\approx& Q_{0}^{(i)}\delta_{I,I'}+Q_{2}^{(i)}\delta_{I\pm2,I'}+\nonumber\\
&&\frac{1}{I}\left[Q_{0}^{(i)}\sqrt{\frac{3}{2}}I_{-}^{b}-Q_{2}^{(i)}I_{+}^{b}\right]\delta_{I+1,I'}+\nonumber\\
&&\frac{1}{I}\left[-Q_{0}^{(i)}\sqrt{\frac{3}{2}}I_{+}^{b}+Q_{2}^{(i)}I_{-}^{b}\right]\delta_{I-1,I'}.
\label{m}
\end{eqnarray}
$I_{+}^{b}$ and $I_{-}^{b}$ are boson realisations of the angular momentum raising and lowering operators corresponding to a certain wobbling phonon number $n$ with an associated wobbling frequency. Note that these are defined in the representation where the projection $K=I$ of their complementary spherical component operator is diagonal. This means that one will further work in a rotated frame of reference with $Q_{0}^{(i)}$ and $Q_{2}^{(i)}$ being the intrinsic quadrupole moments in respect to the rotated frame "$i$" which are related to the commonly used moments associated to the system of reference with axis $3$ as quantization axis by \cite{Varsh}:
\begin{eqnarray}
Q_{0}^{(i)}&=&D_{00}^{2}(\psi_{i},\theta_{i},\phi_{i})Q_{0}+\nonumber\\
&&\left[D_{02}^{2}(\psi_{i},\theta_{i},\phi_{i})+D_{0-2}^{2}(\psi_{i},\theta_{i},\phi_{i})\right]Q_{2},\\
Q_{2}^{(i)}&=&D_{20}^{2}(\psi_{i},\theta_{i},\phi_{i})Q_{0}+\nonumber\\
&&\left[D_{22}^{2}(\psi_{i},\theta_{i},\phi_{i})+D_{2-2}^{2}(\psi_{i},\theta_{i},\phi_{i})\right]Q_{2}.
\end{eqnarray}
Here, symbols $D_{mm'}^{j}$ denote the Wigner functions with their arguments $(\psi,\theta,\phi)_{i}$ being the Euler angles that define the transformation from the original reference frame (with axis 3 as quantization axis) to the rotated frame "$i$" in the $x$-convention \cite{Landau}. The ratio of $Q_{2}$ and $Q_{0}$, is a measure of the deviation from symmetry about the 3 axis, and can be related to parameter $\gamma$ by:
\begin{equation}
\frac{Q_{2}}{Q_{0}}=\frac{\tan{\gamma}}{\sqrt{2}}.\label{tan}
\end{equation}
The angle sequence $(\psi,\theta,\phi)_{i}$ for each case is listed in Table \ref{tab1} while the corresponding relations between $Q$ and $Q^{(i)}$ components are given in Table \ref{tab2}.

\setlength{\tabcolsep}{5.5pt}
\begin{table*}[ht!]
\caption{The expressions of the transformed quadrupole moments $Q_{0}^{(i)}$ and $Q_{2}^{(i)}$ in terms of the original components $Q_{0}$ and $Q_{2}$ which correspond to a reference frame where the third axis is the quantization axis.}
\label{tab2}
\begin{center}
\begin{tabular}{ccc}
\hline\hline\noalign{\smallskip}
$i$ & $Q_{0}^{(i)}$ & $Q_{2}^{(i)}$\\
\noalign{\smallskip}\hline\noalign{\smallskip}
(1) & $-\frac{Q_{0}}{2}+\sqrt{\frac{3}{2}}Q_{2}$ & $-\frac{1}{2}\left(\sqrt{\frac{3}{2}}Q_{0}+Q_{2}\right)$\\
(2) & $-\frac{Q_{0}}{2}+\sqrt{\frac{3}{2}}Q_{2}$ & $-\frac{1}{2}\left(\sqrt{\frac{3}{2}}Q_{0}+Q_{2}\right)e^{2i\alpha_{2}}$\\
(3) & $\frac{1}{4}\left[(1-3\cos{2\alpha_{3}})Q_{0}+2\sqrt{6}\cos^{2}{\alpha_{3}}Q_{2}\right]$ & $\frac{1}{4}\left[-\sqrt{6}\cos^{2}{\alpha_{3}}Q_{0}+(\cos{2\alpha_{3}}-3)Q_{2}\right]$\\
\noalign{\smallskip}\hline\hline
\end{tabular}
\end{center}
\vspace{-0.5cm}
\end{table*}

Performing a power expansion up to the second order of the equations (\ref{Icl}) around the minimum points $(r_{i},\varphi_{i})$, one can then quantize them in the similar way as in the case of the classical energy function. The quantum counterparts of the variables $\tilde{r}_{i}$ and $\tilde{\varphi}_{i}$ can be written in terms of wobbling boson operators as:
\begin{eqnarray}
\tilde{\varphi}_{i}&=&\sqrt{\frac{1}{2I}}\frac{1}{k_{i}}\left(a^{\dagger}+a\right),\nonumber\\
\tilde{r}_{i}&=&i\sqrt{\frac{I}{2}}k_{i}\left(a^{\dagger}-a\right),
\label{oscl1}
\end{eqnarray}
where $k_{i}=\sqrt{m_{i}\omega_{i}}$ with $\omega_{i}$ being the wobbling frequencies corresponding to each minimum point, while
\begin{equation}
m_{i}=\left[I\left(\frac{\partial^{2}{\mathcal{H}}}{\partial{r^{2}}}\right)_{r_{i},\varphi_{i}}\right]^{-1}
\end{equation}
plays the role of oscillator mass. For the sake of completeness, one lists below the explicit expression of mass for each solution:
\begin{eqnarray}
m_{1}&=&\left[(2I-1)(A_{3}-A_{1})+2A_{1}j\right]^{-1},\\
m_{2}&=&\left[(2I-1)(A_{3}-A_{2})\right]^{-1},\\
m_{3}&=&\left[(2I-1)(A_{1}-A_{3})\tan^{2}{\alpha_{3}}\right]^{-1}.
\end{eqnarray}
Using the operator realizations (\ref{oscl1}) in the second order expansions of the classical angular momentum components (\ref{Icl}), one finally obtains the boson representations of the lowering and raising angular momentum operators in the rotated frame
\begin{eqnarray}
I_{+}^{b}&=&\sqrt{\frac{I}{2}}\left[\left(\frac{1}{k_{i}}-k_{i}\right)a^{\dagger}+\left(\frac{1}{k_{i}}+k_{i}\right)a\right],\nonumber\\
I_{-}^{b}&=&\sqrt{\frac{I}{2}}\left[\left(\frac{1}{k_{i}}+k_{i}\right)a^{\dagger}+\left(\frac{1}{k_{i}}-k_{i}\right)a\right].\label{bosr}
\end{eqnarray}
This is one particular boson realization of the angular momentum operators which stems directly from the canonicity of the coordinates $\varphi$ and $r$ and consequently from the correspondence (\ref{oscl1}). It is actually a first order approximation of the Holstein-Primakoff boson expansion \cite{HP} expressed in rotated creation operators. Holstein-Primakoff boson expansion was extensively exploited to describe triaxial nuclei \cite{Tanabe1,Marsh1,Marsh2,Tanabe2,Tanabe3,Tanabe4,Tanabe5}. The full boson expansion can be recovered from the classical functions (\ref{Icl}) by associating a boson algebra to a different pair of canonical coordinates which are complex functions of $\varphi$ and $r$ \cite{RaBu}. Choosing different sets of complex canonical coordinates one arrives at other boson expansions which are commonly used in spin related problems \cite{Dyson,Expo} or new unexplored boson realizations \cite{RaBu}. As all these boson mappings originate from the same classical functions (\ref{Icl}), the later can be considered as the universal classical angular momentum realization.

Plugging the lowering and raising angular momentum operators (\ref{bosr}) into (\ref{m}), one easily obtains the following transition probabilities:
\begin{widetext}
\begin{eqnarray}
B(E2;n,I\,\rightarrow\,n,I\pm2)_{i}&=&\frac{5e^{2}}{16\pi}\left|Q_{2}^{(i)}\right|^{2},\\
B(E2;n,I\,\rightarrow\,n-1,I-1)_{i}&=&\frac{5e^{2}}{16\pi}\frac{n}{2I}\left|Q_{0}^{(i)}\sqrt{\frac{3}{2}}\left(\frac{1}{k_{i}}+k_{i}\right)-Q_{2}^{(i)}\left(\frac{1}{k_{i}}-k_{i}\right)\right|^{2},\label{BE1a}\\
B(E2;n,I\,\rightarrow\,n+1,I-1)_{i}&=&\frac{5e^{2}}{16\pi}\frac{(n+1)}{2I}\left|Q_{0}^{(i)}\sqrt{\frac{3}{2}}\left(\frac{1}{k_{i}}-k_{i}\right)-Q_{2}^{(i)}\left(\frac{1}{k_{i}}+k_{i}\right)\right|^{2}.\label{BE1b}
\end{eqnarray}
\end{widetext}

Another important observable concerning the electromagnetic properties of the wobbling excitations is the $B(M1)$ transition probability. Following the same procedure as in the case of the quadrupole transitions and considering the alignment of the quasiparticle angular momentum $j$ along the first axis \cite{Frau}, one obtains the following expressions for the $B(M1)$ rates connecting different wobbling bands:
\begin{eqnarray}
&&B(M1;n,I\,\rightarrow\,n-1,I-1)_{i}\nonumber\\
&&=\frac{3}{4\pi}\frac{n}{4I}\left|j(g_{j}-g_{R})\left(\frac{1}{k_{i}}+k_{i}\right)\right|^{2},\\
&&B(M1;n,I\,\rightarrow\,n+1,I+1)_{i}\nonumber\\
&&=\frac{3}{4\pi}\frac{(n+1)}{4I}\left|j(g_{j}-g_{R})\left(\frac{1}{k_{i}}-k_{i}\right)\right|^{2}.
\end{eqnarray}
$g_{R}$ and $g_{j}$ are the gyromagnetic factors of the collective core and respectively of the odd particle.

\setcounter{equation}{0}
\renewcommand{\theequation}{3.\arabic{equation}}
\section{Numerical results}
\subsection{Wobbling phase diagram}

\begin{figure}[t!]
\begin{center}
\includegraphics[width=0.45\textwidth]{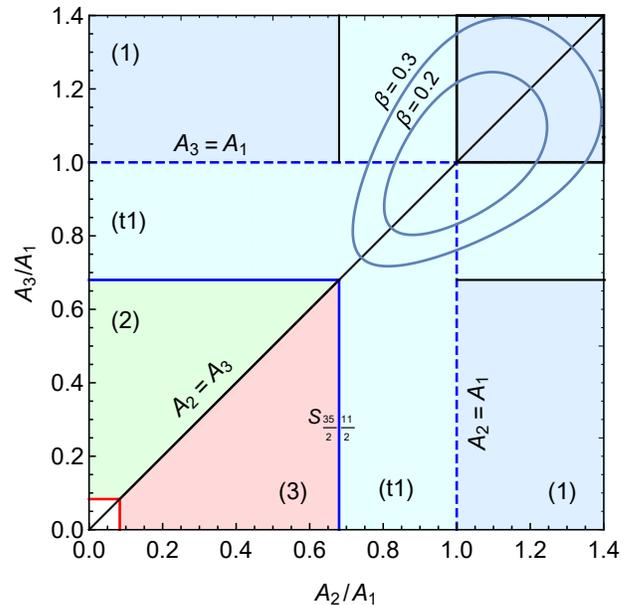}
\end{center}
\caption{Wobbling phase diagram for independent rigid MOI represented for $I=35/2$ and $j=11/2$. In the region near the origin bounded by $S_{13/2,11/2}$ the wobbling modes do not exist. The transversal regime of the first wobbling mode is denoted with $(t1)$. The two closed curves show the relationship among the rigid MOI parametrized as in (\ref{rig}) for $\beta=0.2$ and $\beta=0.3$, when the asymmetry $\gamma_{rig}$ is varied.}
\label{phaserig}
\end{figure}

\begin{figure}[t!]
\begin{center}
\includegraphics[width=0.45\textwidth]{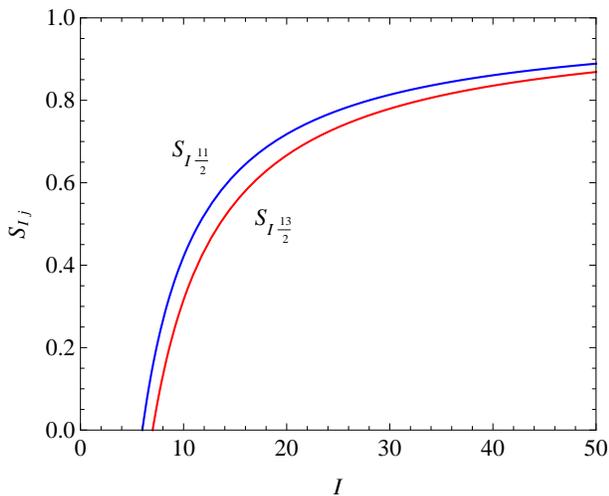}
\end{center}
\caption{Evolution of the separatrix $S_{Ij}$ as a function of angular momentum for $j=11/2$ and $13/2$.}
\label{SIj}
\end{figure}

The complex motion of the resulted wobbling modes and their domains of existence defined in Table \ref{tab1} can be schematically represented as a phase diagram. The three inertia $A_{k}(k=1,2,3)$ can be reduced to only two independent ones, by extracting for example $A_{1}$ as a scaling parameter. In this way one can visualize the wobbling phases as a function of only $A_{2}$ and $A_{3}$ given in units of $A_{1}$. This is done in Fig. \ref{phaserig} for $I=35/2$ and $j=11/2$. The entire phase space is covered by all three wobbling regimes separated by so called separatrices, which are phase space curves where the wobbling vanishes. There is an exclusion region near origin, which defines a minimal angular momentum value where the wobbling excitations can occur. Analytically, this is explained by the fact that for values lower than this limiting spin, the $S_{Ij}$ quantity becomes negative and leads to an imaginary wobbling frequency. The phases corresponding to wobbling modes 2 and 3 are delimited by the $A_{2}=A_{3}$ separatrix. The separatrices between the first wobbling mode and those corresponding to modes 2 and 3 are defined by $S_{Ij}$ which also imply that $|\cos{\alpha_{2,3}}|=1$. This separatrix depends on angular momentum as in Fig. \ref{SIj}. As angular momentum increases, it moves toward its boundary limit where $A_{2,3}=A_{1}$. The first wobbling mode can lead to both increasing and decreasing wobbling frequencies as function of angular momentum \cite{Frau}. Analysing the frequencies (\ref{w1})-(\ref{w3}) for each wobbling mode, one arrives at the conclusion that the first wobbling mode is the only one which can produce decreasing wobbling frequencies. This wobbling regime is called \emph{transversal}, while the remaining mode of the first wobbling phase is called \emph{longitudinal}. The region between $S_{Ij}$ and the limit of $A_{2,3}=A_{1}$ defines a part of the transversal mode of the first wobbling regime where both $A_{2}$ and $A_{3}$ are smaller than $A_{1}$. Such a configuration is supposed to occur when the aligned quasiparticle is of the hole type \cite{Frau}. The second part of the transversal wobbling mode is extended to regions where $A_{2}\lessgtr A_{1}$ and $A_{3}\gtrless A_{1}$. The whole domain of existence for the transversal wobbling mode is shrinking as angular momentum increases. From the phase diagram of Fig. \ref{phaserig} one can observe also that the longitudinal regime of the first wobbling phase is completely separated from the tilted-axis wobbling phases which are bounded only by transversal solutions.

A common parametrization of the rigid MOI is
\begin{equation}
\mathcal{J}_{k}^{rig}=\mathcal{J}_{0}^{rig}\left[1-\beta\sqrt{\frac{5}{4\pi}}\cos{\left(\gamma_{rig}-\frac{2}{3}k\pi\right)}\right],
\label{rig}
\end{equation}
where $\beta$ is the static quadrupole deformation. The evolution of the ratios between these MOI as $\gamma_{rig}$ is varied can be ascertained from the same phase diagram of Fig. \ref{phaserig}. Within the parametrization (\ref{rig}), the tilted-axis wobbling is allowed starting only from a certain angular momentum. This critical value of the angular momentum is lower for more deformed nuclear shapes, {\it i.e.} with a larger $\beta$ deformation. Another notable observation regarding the curves shown in Fig. \ref{phaserig} for ratios of MOI given by Eq.(\ref{rig}) is that they reside predominantly in the existence domain of the transversal wobbling regime.

\begin{figure}[t!]
\begin{center}
\includegraphics[width=0.45\textwidth]{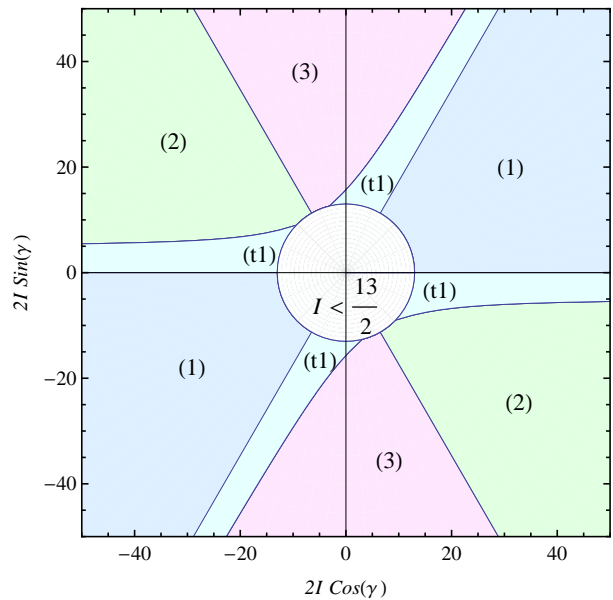}
\end{center}
\caption{Wobbling phase diagram for hydrodynamic MOI represented in the cartesian coordinates $x=2I\cos{\gamma}$ and $y=2I\sin{\gamma}$. In the middle exclusion region the wobbling modes not exist. The transversal regime of the first wobbling mode is denoted with $(t1)$.}
\label{region}
\end{figure}

In order to study the dynamics of the system in each of the wobbling phases, {\it i.e.} the evolution of the phase diagram as a function of angular momentum, one considers for simplicity the hydrodynamic estimation of MOI given by Bohr and Mottelson \cite{BM}:
\begin{equation}
\mathcal{J}_{k}=\frac{4}{3}\mathcal{J}_{0}\sin^{2}{\left(\gamma-\frac{2}{3}k\pi\right)},
\label{inert}
\end{equation}
which is parametrized just by the asymmetry measure $\gamma$ and a scale $\mathcal{J}_{0}$. Using parametrization (\ref{inert}) on the minimum conditions listed in Table \ref{tab1}, one can represent graphically the wobbling phase space with an imbedded angular momentum dependence as in Fig.\ref{region}. The first observation is that the phase diagram has a reflection symmetry in the $\gamma$ shape variable. In the alternative Lund convention \cite{Lund}, both rigid and hydrodynamic MOI are parametrized in the same way as in Eq.(\ref{rig}) and (\ref{inert}) but with an opposite sign for $\gamma$ and respectively $\gamma_{rig}$. This is the reason why the (t1) solution in the fourth quadrant from Fig.\ref{region} was called in previous studies as wobbling in the positive-gamma rotation.

The conclusions made in the analysis with rigid MOI are standing. In this case however, the separatrix between modes 2 and 3 can be expressed by $\gamma=2\pi/2(Mod\,\pi)$, while the interval of existence for the longitudinal mode of the first wobbling phase is given as $\gamma\in(0,\pi/3)(Mod\,\pi)$.  Parametrization (\ref{inert}) reduces the number of parameters at the cost of constricting the domain of values for the moments of inertia. This for example affects very much the existence interval of transversal wobbling with $A_{1}>A_{3},A_{2}$. This condition is fulfilled with hydrodynamic MOI only for $I=13/2$ and $I=15/2$, provided the restrictions
\begin{eqnarray}
90^{\circ}<&\gamma&<103.71^{\circ}\,(Mod\,\pi),\nonumber\\
136.29^{\circ}<&\gamma&<150^{\circ}\,(Mod\,\pi),
\end{eqnarray}
and respectively
\begin{eqnarray}
90^{\circ}<&\gamma&<92.45^{\circ}\,(Mod\,\pi),\nonumber\\
147.55^{\circ}<&\gamma&<150^{\circ}\,(Mod\,\pi),
\end{eqnarray}
are satisfied. Contrary to the case with rigid MOI, these intervals are very narrow. Therefore the configuration of a hole aligned to a triaxial core with hydrodynamic MOI will most probably lead to a tilted-axis wobbling described by the second or third mode. And the wobbling excitations will be an increasing function of angular momentum.

\begin{figure}[t!]
\begin{center}
\includegraphics[width=0.45\textwidth]{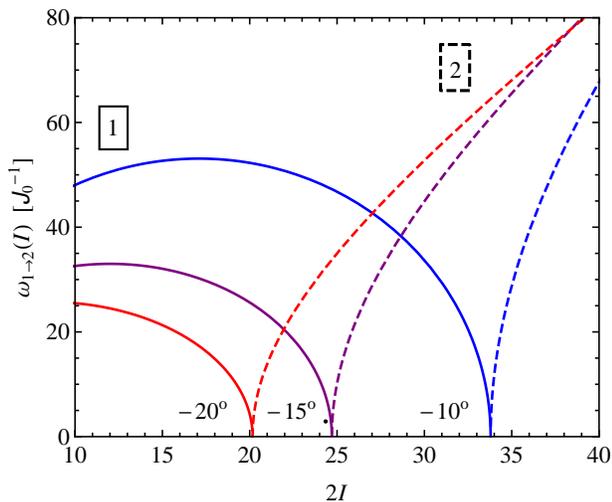}
\end{center}
\caption{Evolution of wobbling frequency given in units of $\mathcal{J}_{0}^{-1}$ as a function of $2I$ from the transversal regime of the wobbling mode 1 to the tilted-axis wobbling mode 2. The curves correspond to $\gamma=-10^{\circ},-15^{\circ}$ and $-20^{\circ}$ with associated critical spins around $I=21/2,25/2$ and $33/2$.}
\label{wtrans}
\end{figure}

Due to the angular momentum dependence of separatrices between the principal axis and tilted-axis wobbling modes, a transition between them is possible when a fixed value of $\gamma$ is considered. As a consequence, the termination of the transversal wobbling band suggested in Refs. \cite{Frau,Chen3} is actually a transition to a tilted-axis wobbling mode \cite{Heiss,Matsu}. The evolution of the wobbling frequencies along such a transition is given in Fig.\ref{wtrans} for few values of the $\gamma$ deformation which connect wobbling modes 1 and 2 in the hydrodynamic parametrization of MOI. The picture for the transition between wobbling phases 1 and 3 is the same, but is obtained for $\gamma$ values shifted with $2\pi/3(Mod\,\pi)$, which amounts to the interchange between hydrodynamic MOI corresponding to second and third principal axes. The decreasing behavior of the transversal wobbling frequency seems to follow an ellipsoidal curvature when hydrodynamic MOI are employed.

\begin{figure}[ht!]
\begin{center}
\includegraphics[width=0.45\textwidth]{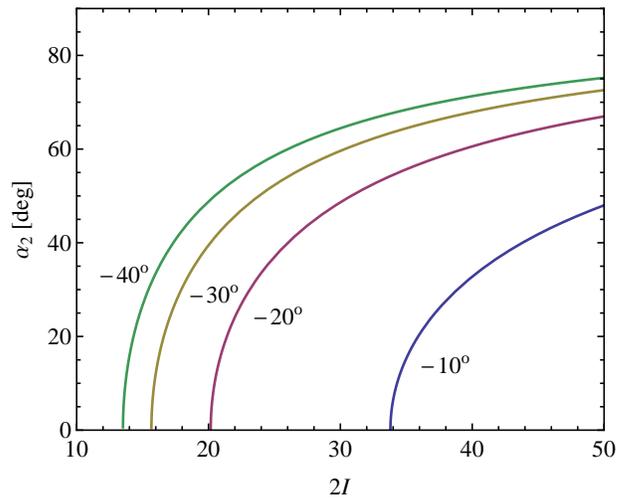}
\end{center}
\caption{Tilting angle $\alpha_{2}$ defined by (\ref{alpha}) is given as a function of $2I$ for few $\gamma$ values from the phase space of the wobbling mode 2 with hydrodynamic MOI.}
\label{a23}
\end{figure}

In Fig. \ref{a23} one plotted the dependence of angle $\alpha_{2}$ on angular momentum for different values of triaxiality $\gamma$. The result is easily transposable to the solution 3 described by the tilting angle $\alpha_{3}$. Thus, in contradistinction to the wobbling mode 1, where the average orientation of the angular momentum vector is constantly along the first principal axis, the mean rotation axis for the rest of the wobbling modes moves from the first principal axis as the spin increases. For asymptotically high angular momentum values, the average rotational axis tends to align itself perpendicularly to the alignment axis 1 $(\sin{\alpha_{2,3}}\rightarrow1)$. But, as can be seen from Fig. \ref{a23}, this full alignment cannot be achieved by experimentally measured high spins states which reach up to $I=97/2$ in TSD bands of few Lu isotopes.  The energy in this case will recover the simple formula of Bohr and Mottelson for wobbling excitations around axis 2 or 3. From Fig. \ref{a23} one can also see that $\alpha_{2}$ increases very rapidly for the first few angular momentum states and then reaches a relative saturation plateau. This means that for sufficiently high starting spin value, an extended part of the wobbling band in this regime can be described by a near constant tilting angle. If $\gamma$ is closer to the $-\pi/3$ separatrix, the increase of $\alpha_{2}$ becomes more abrupt and the plateau more level.

\subsection{Comparison with experiment}

The introduction of the wobbling mode 1 in \cite{Frau,Shimizut}, and especially its transverse regime was a great step into the understanding of the experimentally observed wobbling excitations in Lu and Ta isotopes \cite{Ta}. In these nuclei the wobbling bands are populated up to very high spin states and the data do not exhibit any trace of band termination. This means that it happens at a very high spin, which is not yet reached by experiment. The $^{135}$Pr nucleus is a little different from all other measured wobbling excitations. First of all it is based on an aligned proton from the $h_{11/2}$ orbital instead of $i_{13/2}$ as it happens in Lu and Ta. Secondly, it exhibits a backbending-like anomaly in its yrast and wobbling bands. This anomaly results in an inversion of the angular momentum dependence of the wobbling energy which is defined as:
\begin{eqnarray}
E_{W}^{(i)}&=&E_{i}(I,1)-\frac{1}{2}\left[E_{i}(I-1,0)+E_{i}(I+1,0)\right]\nonumber\\
&=&\frac{1}{2}\left\{3\omega_{i}(I)-\frac{1}{2}\left[\omega_{i}(I-1)+\omega_{i}(I+1)\right]\right\}-\delta E_{i},\nonumber\\
\label{Ew}
\end{eqnarray}
where $i=1,3$ and
\begin{eqnarray}
\delta E_{1}&=&A_{1},\\
\delta E_{2}&=&A_{2}-\frac{2A_{1}j\cos{\alpha_{2}}}{(2I-3)(2I+1)},\\
\delta E_{3}&=&A_{3}-\frac{2A_{1}j\cos{\alpha_{3}}}{(2I-3)(2I+1)}.
\end{eqnarray}

\begin{figure}[t!]
\begin{center}
\includegraphics[width=0.45\textwidth]{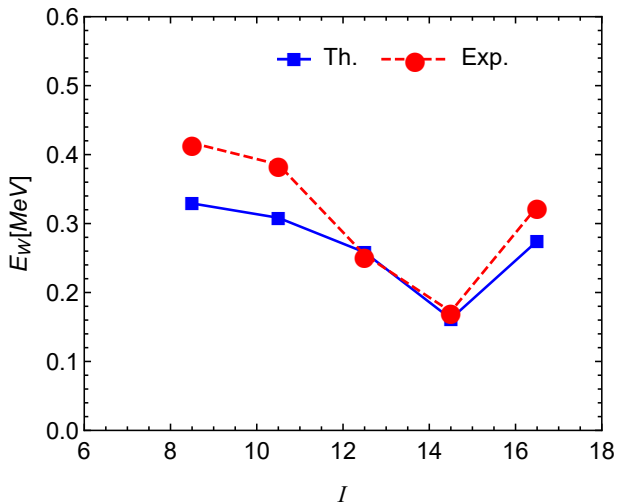}
\end{center}
\caption{Experimental excitation energies \cite{Matta} of the wobbling band in $^{135}$Pr compared with the theoretical wobbling energy.}
\label{wobex}
\end{figure}

Judging by the experimental energy spectrum and measured electromagnetic transitions \cite{Matta}, the transversal mode of wobbling and its associated rotational regime is preserved up to $I=27/2$ in the yrast band ($n=0$) and up to $29/2$ in the wobbling band ($n=1$). The following $I=31/2$ and $I=35/2$ yrast states and the wobbling state $I=33/2$ are then considered to be part of a different rotation-wobbling regime. The low spin at which this transition takes place, excludes the non-independent rigid MOI description (\ref{rig}) of the nuclear shape. Indeed, the small quadrupole deformation of $^{135}$Pr ($\beta\approx 0.18$) \cite{Matta} implies a very high critical angular momentum where the tilted-axis wobbling with non-independent rigid MOI can be achieved. The transition between different wobbling regimes at such a low spin is however possible within the hydrodynamic parametrization of the MOI. Using $I=29/2$ and $I=31/2$ as spins for transversal wobbling termination, one can obtain an interval of $\gamma$ deformations which might describe the transition in the experimental wobbling excitation energy visualized in Fig. \ref{wobex}.  Thus, one obtains two limiting curves of the types shown in Fig. \ref{wtrans} corresponding to $\gamma=-12.13^{\circ}$ and $\gamma=-11.16^{\circ}$.

In the present model, one considers that $\mathcal{J}_{0}$ and $\gamma$ are free parameters which are fixed by fits to experimental data. Fitting just the wobbling energy with Eq.(\ref{Ew}), one can obtain very good agreement with experiment. Such a result will however be far from experimental data in what concerns the angular momentum evolution of the energy spectrum in each considered band. Therefore, the fitting must be performed on the experimental energy levels. To improve the agreement with experiment, the parameter $\mathcal{J}_{0}$ is often amended with a spin dependence \cite{Matta,Chen3,Tanabe3,Tanabe4,Tanabe5}. Here, one will use a different approach inspired from the observations regarding the rotation-vibration collective states. There is a long standing debate wether the lowest excited $K^{\pi}=0^{+}$ band can be interpreted as a rotational band constructed on a $\beta$ vibration excitation \cite{Garret}. The doubt about this interpretation is mostly connected to the persistent failure of the geometrical and algebraical models to reproduce the correct level spacing in these bands while describing the ground and $\gamma$ bands with a high precision. This inconsistency originates from the use of the same inertia for vibrations and rotations. This restriction can be bypassed for example by considering an energy dependent collective potential \cite{Caprio,Noi}.

\begin{figure}[t!]
\begin{center}
\includegraphics[width=0.45\textwidth]{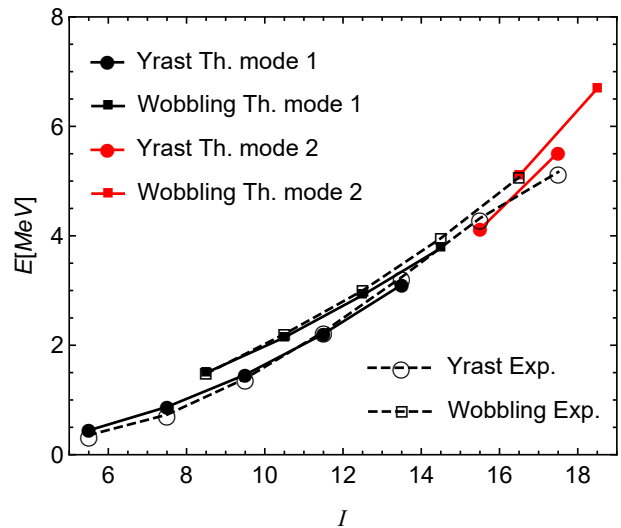}
\end{center}
\caption{Comparison of $n=0$ and $n=1$ energy levels between theoretical results and experimental data \cite{Matta} for $^{135}$Pr.}
\label{ebands}
\end{figure}

The dynamical distinction between the vibrational and rotational degrees of freedom will be used here for fitting the experimental energy levels of the yrast band and wobbling band of $^{135}$Pr with two different scaling parameters, $\mathcal{J}_{0}^{R}$ for the rotation-precession terms and $\mathcal{J}_{0}^{W}$ for the wobbling frequency. The parameters resulted from the fits are $\mathcal{J}_{0}^{R}=30.96$ MeV$^{-1}$, $\mathcal{J}_{0}^{W}=65.93$ MeV$^{-1}$ and $\gamma=-11.18^{\circ}$, which corresponds to an rms=0.149 MeV. The comparison between the theoretical and considered experimental energy levels is made in Fig. \ref{ebands}. As can be seen, the agreement with experiment is quite good, even in the transition region between $I=27/2$ and $I=33/2$. This is also reflected in an impressive reproduction of the wobbling energy evolution as a function of spin shown in Fig. \ref{wobex}. The triaxial deformation $\gamma=-11.18^{\circ}$ gives the following ratio between the MOI $\mathcal{J}_{1}:\mathcal{J}_{2}:\mathcal{J}_{3}=15:24:1$. The transversal character of the wobbling excitations up to $I=29/2$ is evident from the fact that the MOI of the principal axis around which the rotation takes place has the intermediate value \cite{Frau}. The tilted-axis wobbling corresponds to a rotation axis which departs from the first principal axis toward the second principal axis. The angle $\alpha_{2}$ between the rotation axis of the transverse wobbling and that of the tilted-axis mode 2 is 4.22$^{\circ}$, 20.78$^{\circ}$, and 28.36$^{\circ}$ for the states $I=31/2,33/2$ and respectively $35/2$.

The fitted parameters are used to calculate transition probabilities, which are compared with few experimental values in Table \ref{tab3}. The ratio between electric quadrupole transition is scale free, while the ratio between the inter-band $M1$ transition and the in-band $E2$ transition have a scale dependence on the quantity $[(g_{j}-g_{R})/Q_{0}]^{2}$. Due to the lack of information regarding $Q_{0}$ and the uncertainty in the degree of quenching for the gyromagnetic factor, the theoretical calculations employ for this ratio the value of 0.0502 which equates the experimental data for $I=25/2$ state. The theoretical results for $E2$ transitions are in the range of experimental data. The experimental values are however decreasing more rapidly with spin than the theoretical ones. At the termination spin $I=29/2$ the theoretical calculations show an increase. This  behavior is not excluded by experimental data, but for confirmation one needs more precise measurements. The theoretical magnetic transition ratios also exhibit a decreasing with spin, which is in contradiction with the state dependence of the experimental data. Once again the transition probability is enhanced for the terminating spin $I=29/2$.

\setlength{\tabcolsep}{5.5pt}
\begin{table}[ht!]
\caption{Experimental \cite{Matta} and theoretical $E2$ and $M1$ transition probabilities for transitions from the $n=1$ wobbling band to $n=0$ band. The rates are normalized to the E2 transition within the wobbling band from the same state.}
\label{tab3}
\begin{center}
\begin{tabular}{ccccc}
\hline\hline\noalign{\smallskip}
 & \multicolumn{2}{c}{$\frac{B(E2,I\rightarrow I-1)_{out}}{B(E2,I\rightarrow I-2)_{in}}$} & \multicolumn{2}{c}{$\frac{B(M1,I\rightarrow I-1)_{out}}{B(E2,I\rightarrow I-2)_{in}}\,\left(\frac{\mu_{N}}{eb}\right)^{2}$}\\
\noalign{\smallskip}\cline{2-5}\noalign{\smallskip}
$I$ & Exp. & Th. & Exp. & Th.\\
\noalign{\smallskip}\hline\noalign{\smallskip}
$\frac{17}{2}$ & & 0.313& &0.164\\
\noalign{\smallskip}
$\frac{21}{2}$ &0.843(32) & 0.270 & 0.164(14) &0.164\\
\noalign{\smallskip}
$\frac{25}{2}$ &0.500(25) & 0.258& 0.035(9)&0.183\\
\noalign{\smallskip}
$\frac{29}{2}$ &$\geq$0.261(14) &0.318 &$\geq$0.016(4) &0.279\\
\noalign{\smallskip}\hline\hline
\end{tabular}
\end{center}
\vspace{-0.5cm}
\end{table}

One expects an improvement of the agreement with experimental energy levels and especially transition probabilities, for fits against independent rigid MOI. This however increases the number of free parameters. Alternatively, a better reproduction of data can be achieved by coupling the wobbling excitations to scissor-like oscillations of the proton and neutron distributions \cite{Frau1}.

\section{Conclusions}

Three unique wobbling phases are determined as quantized oscillations around minima in the classical energy associated to a quantum triaxial rotor Hamiltonian with an aligned single-particle angular momentum along the first principal axis, by means of a time-dependent variational principle. From the dynamical point of view, the three phases correspond to two distinct pictures: The first phase describes the single-particle angular momentum alignment along the first principal axis which is the approximate rotation axis. It gathers both longitudinal and transverse regimes discussed in Ref.\cite{Frau}. The other two wobbling phases describe tilted-axis wobbling excitations, with the approximate rotation axis contained in the principal planes defined by the first principal axis with the second and respectively the third axes. The existence conditions for each wobbling mode are discussed in terms of independent MOI. All three wobbling phases are bordered by separatrices. One distinguished two kinds of separatrices, dependent and independent on spin. The first type, could be crossed during the increase of angular momentum. Such a situation corresponds to a wobbling phase transition between the first phase and any other tilted-axis wobbling modes. Using the hydrodynamic parametrization of the MOI, a phase diagram was drawn in terms of the triaxiality parameter $\gamma$ and the total angular momentum $I$. In such a phase diagram, the transition paths with stable $\gamma$ deformation can start only from the transversal subspace of the first wobbling phase. The transition from the transversal wobbling to a tilted-axis regime is used to describe the wobbling excitations in $^{135}$Pr nucleus. The agreement with experiment is very good in what concerns the energy levels. The theoretical results are able to reproduce the discontinuity in the wobbling energy ascribed to the aforementioned transition. The results of the fits were used to calculate $E2$ and $M1$ transition probabilities connecting the wobbling band states with the yrast energy levels. The domain of values is close to the experimental data on electric transitions. However the reproduction of the spin evolution is deficient, especially for magnetic transitions. An enhancement of the transitions connecting the last transverse wobbling state $I=29/2$ is observed in the theoretical calculations. This is due to the proximity of the wobbling separatrix, which makes the wobbling solutions unstable.

In conclusion, besides reproducing the known result for transverse and longitudinal wobbling, one completed the wobbling phase space with a tilted-axis wobbling mode. In this way, the whole dynamical description of the particular system composed of a triaxial core and an aligned quasiparticle is treated in a unified manner. Each wobbling mode follows strict conditions associated to the MOI. This analysis put some additional constraints to the transverse wobbling regime introduced in Ref. \cite{Frau}. Nevertheless, it is still experimentally realizable such that its existence and stability cannot be disputed \cite{Tanabe5,Frau2}.

\section*{Acknowledgments}
Support of CNCS-UEFISCDI, through project number PN-III-P4-ID-PCE-2016-0092 is kindly acknowledged.

\end{document}